\documentclass[pre,aps,preprint,showpacs,groupedaddress]{revtex4}

    \usepackage[psamsfonts]{amsfonts,amssymb}
    \usepackage{bm}
    \usepackage{graphicx}
    \usepackage[dvips]{color} 
    \usepackage[pdfmark,colorlinks]{hyperref} 

\newcommand{\mpArc}[1]{
           {\tt \href{http://www.ma.utexas.edu/mp_arc-bin/mpa?yn=#1}
           {\goodbreak mp\_arc~#1}}}
\newcommand{\arXiv}[1]{{\tt \href{http://arXiv.org/abs/#1}
           {\goodbreak #1}}}

\begin{document}
                    \title{
Variational method for locating invariant tori
                        }
                    \author{
Y. Lan$^1$, C. Chandre$^2$ and P. Cvitanovi\'{c}$^1$
                        }
        \affiliation{ $^1$
Center for Nonlinear Science,
School of Physics, Georgia Institute of Technology,
Atlanta, GA 30332-0430
    \\ $^2$
Centre de Physique Th\'eorique,
Luminy, Case 907,
F-13288 Marseille cedex 09, France
                     }

        \date{\today}
        \begin{abstract}
We formulate a variational fictitious-time flow
which drives an initial guess torus to a torus invariant
under given dynamics.
The method is general and applies in principle to continuous time flows and
discrete time maps in arbitrary dimension,  and to both Hamiltonian
and dissipative systems.
        \end{abstract}
        \maketitle

\section{Introduction}

    Analysis of dynamical systems in terms of invariant phase space
structures provides important insights into {the} behavior of physical systems.
The simplest such invariants are equilibria, points in phase space
which are stationary solutions
  or 0-dimensional invariants
of the flow. They and their stable/unstable manifolds yield information
about the topology of the flow. The role that the next class of flow invariants,
periodic orbits, play in {the} topological organization of phase space and
{the} computation of long time chaotic dynamics averages is well known (for
an overview, see Ref.~\cite{Bchaos}).  A periodic orbit is topologically
a circle or an invariant 1-torus for a flow, and a set of discrete points or an
invariant 0-torus for a map, embedded in a $d$-dimensional phase
space.
Higher dimensional invariant tori also frequently
play an important role in the dynamics; we refer the reader to
Ref.~\cite{WysMeis05} for some of the references to this literature.
Invariant tori of dimension lower than the dimension of the dynamical flow
can be normally hyperbolic~\cite{Treshev94,hir77inv}.
KAM theory  implies that invariant tori occur in Cantor sets, and
such tori play key roles in the phase-space transport~\cite{mac84tr,mac84st}.
For 2-degree of freedom Hamiltonian flows (i.e., 4-dimensional phase space),
2-dimensional invariant tori act as barriers
 to diffusion through phase space, and for higher dimensional flows
such structures act as effective barriers (Arnold diffusion). In dissipative
systems such as  Newtonian fluids, quasi-periodic motion on two or higher dimensional
tori is one of the routes to the eventual turbulent motion~\cite{cvt89b,nhouse78}.

There are many methods for determining periodic orbits
available in the literature~\cite{Bchaos,chen87orb,mes87new}.
The lack of comparably effective methods for the determination of higher dimensional
invariant structures has stymied the exploration of {the}
phase spaces of high-dimensional flows, a focus of much recent
research~\cite{WysMeis05,delaLlave04:348}.

Signal processing methods like frequency analysis~\cite{laskar,lask92},
based on the analysis of trajectories,
can detect elliptic invariant tori since these tori
influence significantly the behavior of nearby trajectories.
Bailout methods~\cite{cart02bail,bab00dyn}
can effectively locate the elliptic regions in a non-integrable system, 
by embedding the dynamical system into a 
larger phase space. Ref.~\cite{bot95find} describes a
variational technique designed to find regular orbits in a phase
space with mixed dynamics. However, these methods can only
detect trajectories with non-positive Lyapunov exponents. They single out 
regular motions in a phase space but can not exactly determine a 
torus unless it is stable. Due to their relative ease of identification,
in particular cases periodic orbits are used to study invariant tori and
their breakup. For example,
in Greene's criterion approach~\cite{gree79,mack92,tom96num} one studies a
sequence of periodic orbits which converges to a given invariant torus.
Such approaches have been mainly applied to the
determination of tori of Hamiltonian systems with 2 degrees of freedom.

Other techniques to determine invariant tori are
specific to the phase space dynamics of the system under consideration, 
most often a Hamiltonian system. Early attempts like spectral balance 
method were based on the computation
of quasi-periodic orbits~\cite{park89num,Simo90}, the closure of which constitutes the invariant 
torus. To overcome the small divisor problems associated with the flow on 
a torus, recent research employed a geometric point of view and focused on
the invariant torus itself. Efforts are devoted to find the solution of the
so-called invariance condition which ensures the invariance of a parametrized 
object in phase space. Invariance conditions are functional
equations for maps~\cite{kev85num,deb94num,moo96com,vel87new,br97alg,die91num} 
and first-order partial differential equations (PDEs) for flows~\cite{min97com,ge98con,sch05con}.
These equations can be solved by Newton's method or Hadamard graph transform 
technique~\cite{hir77inv}. In view of the periodicity in the angle variables, 
Fourier transforms are widely used in the computation%
~\cite{lop05rel,war87inv,CasJorb00,jor03inv,sh05fou,kaa95con,war91clo}. For
Hamiltonian systems, the action principle and the
Hamilton-Jacobi equation are also frequently used in the calculation of periodic
and quasi-periodic orbits~\cite{per74var,per79var,kook89,war87inv,Mather91,gab92iter}.

In this paper we introduce a  method to solve the invariance condition
equation and  obtain invariant {$m$-tori of flows and maps}
embedded in $d$-dimensional phase spaces. It has a variational structure which
guarantees the global convergence.
The method is a
generalization of the variational ``Newton descent'' method originally developed
to locate periodic orbits of flows~\cite{CvitLanCrete02,lanVar1},  which
can be viewed as
a variant of multi-shooting method in boundary value 
problems~\cite{kellbv,bl,nr}. When the representative points on the guess torus
achieve a near-continuous distribution, a PDE is derived which governs their
evolution to a true invariant torus. In spirit, this is similar to the approach
used in Ref.~\cite{war87inv} and thus high accuracy is expected. However, our
method is stable and thus applies to more general cases, including the
searches for partially
hyperbolic tori embedded in the chaotic regions of phase space.
In a general dynamical system, the
phase space structure can be extremely complex,
and the global stability of our algorithm is of key importance for the
the efficiency of our searching program.
In our numerical computation, an adaptive scheme is used which keeps 
changing the step size according to
the smoothness of the evolution.
In addition to the adaptive step size,
we further speed up our searches by utilizing the continuity of the 
variational evolution equations.
These points will be explained in detail in what follows.

In Sec.~\ref{sec:der} we derive the variational equation which governs
the fictitious time dynamics. The numerical implemention of this equation
is discussed in Sec.~\ref{sec:impl}. The method is further illustrated
in Sec.~\ref{sec:num}
through its application to {the determination of 1-tori of} the 
standard map, {of 2-tori of} a forced pendulum {flow (3-dimensional 
phase space)}, {of 1- and 2-tori of} two
coupled standard maps {(a four dimensional symplectic map)},
and {of 2-tori of} the Kuramoto-Sivashinsky system {(infinite dimensional phase space)}.
In particular, we
provide evidence that the method converges up to the threshold of
existence of a given invariant torus and yields
estimates of the critical thresholds {of the
breakup of invariant tori of} 2-degree of freedom
Hamiltonian {systems}.

\section{Newton descent method for invariant tori}
\label{sec:der}

 We start by deriving a variational
 fictitious time evolution equation for {the}
determination of a 1-dimensional invariant torus of a $d$-dimensional map
${\bf f}: \mathbb{R}^d \to \mathbb{R}^d$. The method can be
extended to {the} determination of invariant $m$-tori of
$d$-dimensional maps and flows.

A fixed point (0-dimensional invariant torus) ${\bf x} ={\bf f}({\bf x})$ is a point
which is mapped into itself under {the} action of  ${\bf f}$.
Likewise,
a 1-dimensional invariant torus of ${\bf f}$ is a loop in $\mathbb{R}^d$ which is
mapped into itself under {the} action of  ${\bf f}$. If points
on the invariant 1-torus are parametrized by a cyclic variable $s \in [0,2 \pi]$,
with ${\bf x}(s)={\bf x}(s+2\pi)$, a point ${\bf x}(s)$ is mapped
into another point on the invariant torus
\begin{equation}
 {\bf f}({\bf x}(s)) = {\bf x}(s+\omega(s))
\,,
\label{torusMap}
\end{equation}
where $\omega (s)$ is the local parametrization $s$-dependent
shift. In other words, the full phase space dynamics ${\bf f}$
induces a 1-dimensional circle
map on the invariant 1-torus
\begin{equation}
s \mapsto s+\omega(s) \qquad \mbox{mod}~2\pi
\,.
\label{circMap}
\end{equation}

We also parametrize our guess for the invariant 1-torus,
the loop
${\bf x}(s,\tau)$, by $s \in [0,2\pi]$,
with
${\bf x}(s,\tau)={\bf x}(s+2\pi,\tau)$.
Together with
the ``fictitious time'' $\tau$, to be defined below, this parametrizes a
continuous family of  guess loops.
However, for an arbitrary loop there is no unique definition of the shift
$\omega $, as the loop is not mapped into itself under action of
$ {\bf f}$.
Intuitively, $\omega$ should be fixed by requiring that
the $d$-dimensional distance vector
between the circle map image of a point on the loop at $s$, and
the ``closest'' point on the iterate of the loop
\begin{equation}
{\bf F}(s,\tau)
    =   {\bf x}(s+\omega(s,\tau),\tau)- {\bf f}({\bf x}(s,\tau)),
\label{distMap}
\end{equation}
is minimized.  For example,
if the guess loop is sufficiently close to the desired invariant
1-torus, {$\omega(s,\tau)$} can be fixed by intersecting the loop
with a hyperplane normal to the loop and cutting
through the image of loop {${\bf f}({\bf x}(s,\tau))$}.

Compared with fixed point and periodic orbit searches for iterates of maps,
the new aspect here is that we are searching for $m$-dimensional compact
invariant hyper-surfaces, with  points on such hyper-surfaces
parametrized by $m$ cyclic variables. We have encountered this
situation already for the continuous time flows, for which a periodic orbit $p$
is an invariant 1-torus, with $x(t)  \in p$ naturally parametrized by the cyclic
time variable $t \in [0,T_p]$. For other cyclic coordinates we are
free to choose  a parametrization $s$ that best suits our purposes.

In this exploratory foray into {the} world of compact higher-dimensional invariant
manifolds
we shall make the simplest choice at each turn. In particular, we are free to choose any
parametrization $s$ which preserves ordering of points along the invariant 1-torus,
{\em i.e.} any circle map (\ref{circMap}) that is strictly monotone,
$1+d\omega/ds > 0$. For an irrational rotation number
a strictly monotone circle map can be conjugated to a
constant shift, so in what follows
we {\em define} the $s$ parametrization dynamically,
by  requiring that the action of the dynamics
${\bf f}$
for both the guess loop and the target
invariant 1-torus is rotation with a constant shift $\omega$,
\begin{equation}
s \mapsto s+\omega \qquad \mbox{mod}~2\pi
\,.
\label{shiftMap}
\end{equation}
The invariance condition (\ref{torusMap}) with conjugate dynamics (\ref{shiftMap})
has been used previously in the literature~\cite{CasJorb00,jor03inv}. We now
design a stable scheme which yields a parametrization ${\bf x}(s)$
satisfying Eq.~(\ref{torusMap}) together with Eq.~(\ref{shiftMap}).

Following the approach of Refs.~\cite{CvitLanCrete02,lanVar1}  originally developed
to locate periodic orbits of flows, we now introduce
the simplest cost functional that measures the average distance squared
 (\ref{distMap})  of the guess loop from its iterate
\begin{equation}
\mathcal{F}^2[\tau]
    =  \oint\!\frac{ds}{2\pi}  {\bf F}(s,\tau)^2
\,.
\label{eq:funal}
\end{equation}
Similar functional was used in the stochastic path
extremization~\cite{Onsager53}. Here
$\mathcal{F}^2[\tau]=\mathcal{F}^2[{\bf x},\omega]$
is a functional, as it depends on
the infinity of the points
${\bf x}(s,\tau)$
that constitute the loop {for a given $\tau$}.
If the loop is an invariant 1-torus, $\mathcal{F}^2=0$, otherwise
$\mathcal{F}^2>0$.
At fictitious time
$\tau$ we compute cost due to the two mappings: one is the
iterate ${\bf f}({\bf x}(s,\tau))$ of the loop, and the other  the circle map
$s \mapsto s+\omega(\tau)$ along the loop.
The fictitious time evolution should  monotonically decrease
the distance between a loop and its iterate, as measured by the functional
$\mathcal{F}^2[\tau]$,
by moving  both the totality of loop points ${\bf x}(s,\tau)$ and
modifying the shift $\omega(\tau)$.

With constant shift circle map (\ref{shiftMap})
the variation of $\mathcal{F}^2[\tau]$ under the (yet unspecified)
fictitious time variation $d\tau$ is
\begin{equation}
\frac{d ~} {d\tau} \mathcal{F}^2[\tau] \,=\,
      2 \, \oint\!\frac{ds}{2\pi}
    \left(
    {\bf F} (s,\tau) \cdot \frac{d {\bf F}} {d\tau} (s,\tau)
    \right)
\,,
\label{eq:fvnal}
\end{equation}
where
\begin{eqnarray}
\frac{d ~} {d\tau} {\bf F}(s,\tau) &= &
    \frac{\partial {\bf x}}{\partial \tau}(s+\omega(\tau),\tau)
                \,+\,  {\bf v}(s+\omega(\tau),\tau) \frac{d \omega(\tau) }{d \tau~}
        \nonumber\\ & & \qquad\qquad
        \,-\,  J({\bf x}(s,\tau)) \frac{\partial {\bf x}}{\partial \tau}(s,\tau)
 \, ,
\nonumber\\
 {\bf v}(s,\tau) &= &
    \frac{\partial {\bf x}}{\partial s}(s,\tau)
\,.
\nonumber
\end{eqnarray}
The adjustment in the loop tangent direction
${\bf v}$
is needed to redistribute  points along
the loop in order to ensure the constant shift parametrization $s$,
and the $[d\!\times\!d]$  Jacobian matrix of the map
$J = {\partial {\bf f}}/{\partial {\bf x}}$ {moves} the
loop point ${\bf x}(s,\tau)$  in the ``Newton descent" direction.

Again we design a fictitious time flow in the space of loops
by taking the simplest choice, in the spirit of the Newton method:
\begin{equation}
\frac{d \, {\bf F}} {d\tau}  = -  {\bf F}
\,,
\label{eq:fvch}
\end{equation}
for which $\mathcal{F}^2[{\bf x},\omega]$ decreases
exponentially with fictitious time $\tau$:
 \begin{equation}
\mathcal{F}^2[\tau]
 =\mathcal{F}^2[0] e^{-2 \tau}
\,.\label{eq:fvdc}
\end{equation}
The ``Newton descent'' PDE (\ref{eq:fvch}) which evolves loop points in
fictitious time $\tau$ and along loop direction  $s$  is the main result of this paper.
Written out in detail, the Newton descent equation for a guess loop,
\begin{eqnarray}
&&
\frac{\partial {\bf x}} {\partial \tau} (s+\omega,\tau)
 +\frac{\partial {\bf x}}{\partial s}(s+\omega,\tau)
    \frac{\partial \omega}{\partial \tau}(\tau)
        \label{eq:vtorus}
        \\ && \;\;\;
    -J({\bf x}(s,\tau))  \frac{\partial {\bf x}}{\partial \tau}(s,\tau)
        \,=\, {\bf f}({\bf x}(s,\tau))-{\bf x}(s+\omega,\tau)
\,,
\nonumber
\end{eqnarray}
evolves points ${\bf x}(s,0)$ on the $\tau =0$  initial guess loop
to the points ${\bf x}(s)={\bf x}(s,\infty)$,
$s \mapsto s+\omega$, $\omega = \omega(s,\infty)$,
 on the target 1-torus, provided
that the $\tau$ flow does not get trapped in a local minimum with
$\mathcal{F}^2[\infty]>0$.

The Eq.~(\ref{eq:vtorus}) can also be derived via a multi-shooting
argument as has been done in Ref.~\cite{lanVar1}. Instead of a blind minimization of the 
cost functional (\ref{eq:funal}), the method  uses the vector equation 
(\ref{distMap}) and its first derivative to find the zeros of the cost functional. The  momotonicity of
(\ref{eq:fvdc}) with $\tau$ ensures
 the global convergence. A similar argument has been used in
the derivation of a globally convergent modified Newton's method in Ref.~\cite{nr}.

Generalization to searches for invariant $m$-tori is immediate:
the guess  $m$-torus is parametrized by  
${\bf s}=(s_1,s_2,\ldots,s_m)\in [0,2\pi]^m$, periodic in each cyclic coordinate
\begin{equation}
{\bf x}({\bf s}+2\pi{\bf k})={\bf x}({\bf s})
\, \quad \mbox{ for all } {\bf k}\in{\mathbb Z}^m
\,, \label{eq:vpbc}
\end{equation}
with  $m$ incommensurate shifts
${\bm\omega}=(\omega_1,\omega_2,\ldots,\omega_m)$~\cite{arn89m}.
Now the fictitious time flow~(\ref{eq:vtorus}) has an $[d\!\times\!m]$
invariant surface tangent tensor ${\bf v}$.
The fictitious time flow searches (\ref{eq:vtorus}) for invariant tori can
also be adopted to continuous time flows, by reducing
the flow to  a Poincar\'{e} return map on any local
Poincar\'{e} section which intersects the trajectories on the guess ($m$+1)-torus
transversally.  We will provide examples in what follows.

In general, each tangent vector of an invariant $m$-torus
transformation along given cyclic parameter $s_k$ has a unit
eigenvalue, and requires a constraint.
For example,
for the Jacobian matrix of a continuous time periodic orbit (a 1-torus)
the velocity vector is an
eigenvector with a unit eigenvalue, and Newton descent equations need to be supplemented
with a constraint (a Poincar\'e section) in order to determine the period of the orbit.
In addition,  if the flow is Hamiltonian,
and the invariant $m$-torus
is located on a fixed energy surface $H({\bf p},{\bf q})=E$,
the constraint $d{H}/d\tau=0$ is needed to ensure the conservation of the energy
by the fictitious time dynamics.

In case at hand, there are
two alternative ways to impose the constraint:
We may or may not fix $\omega$ {\em a priori}.

(a)
If we are searching for an invariant 1-torus of
a fixed shift $\omega$,  the
fictitious time flow should not change the shift along the loop,
\begin{equation}
{d\omega}/{d\tau}=0
\,.
\label{eq:fixshift}
\end{equation}.

(b)
If we are
searching for an invariant 1-torus of {a} given topology,
the shift
$\omega=\omega(\tau)$ varies
with the fictitious time $\tau$, and
is to be determined simultaneously with the 1-torus itself.
In this case we impose the {\em phase condition}~\cite{sch05con}
\begin{equation}
\oint ds\!
    \left(
    {\bf v}(s,\tau)  \cdot \frac{\partial {\bf x}}{\partial \tau}(s,\tau)
    \right) =0
\,, \label{eq:nonsp1}
\end{equation}
 which ensures that during the fictitious time evolution
the average motion of the points along the
loop  equals zero. Empirically, for  this  global loop
constraint the fictitious time dynamics is more stable than
for a single-point constraint such as $\delta{\bf x}(0,\tau)=0$.
For $m$-torus, ${\bf v}(s,\tau)$ is a $[d \times m]$ tensor and 
Eq.~(\ref{eq:nonsp1}) yields $m$ constraints. For energy
 conserving Hamiltonian systems, one phase condition has  to be replaced by the
energy condition
\begin{equation}
\frac{1}{2\pi}\oint ds \! \nabla H({\bf x}(s,\tau))\cdot 
\frac{\partial {\bf x}(s,\tau)}{\partial \tau}=E-\frac{1}{2\pi}ds \oint 
H({\bf x}(s,\tau))
\,, \label{eq:encond}
\end{equation}
where a fixed $E$ fixes the energy shell under consideration.

The two cases are
analogous to continuous time
Hamiltonian flow periodic orbit constraints: case (a)
corresponds to fixing the period and varying the energy shell, and case (b) to fixing the energy
and computing the period of a periodic orbit of a given topology.

The examples
of Secs.~\ref{sec:standm}, \ref{sec:forcP} and \ref{sec:2stMap}
illustrate the constant shift $\omega$ constraint
(\ref{eq:fixshift}); the examples
of Fig.~\ref{f:smomg} and Sec.~\ref{sec:KS} illustrate the phase condition
(\ref{eq:nonsp1}).

\section{Numerical implementation}
\label{sec:impl}

Due to the periodic boundary condition
(\ref{shiftMap}) it is convenient to expand the loop point ${\bf x} $,
the Jacobian matrix $J$, the map ${\bf f}$, and the loop tangent
${\bf v}$
as a discrete Fourier series
 \begin{eqnarray}
 {\bf x}(s,\tau) &=& \sum_k
                      {\bf a}_k(\tau) e^{ik s}
        \nonumber\\
       J({\bf x}(s,\tau)) &=&
       \sum_k
                      J_k(\tau) e^{iks}
            \nonumber\\
         {\bf f}({\bf x}(s,\tau)) &=&
       \sum_k
                      {\bf b}_k(\tau) e^{iks}
        \nonumber\\
         {\bf v}(s,\tau) &=&
       i \sum_k
                       k \, {\bf a}_k(\tau) e^{iks}
\label{eq:frx}
 \end{eqnarray}
(${\bf a}_k^*={\bf a}_{-k}$ due to the reality of ${\bf x}(s,\tau)$,
and similar relations hold for $J_k$ and ${\bf b}_k$),
and rewrite
the Newton descent PDE (\ref{eq:vtorus})  as
an infinite ladder of ordinary differential equations:
\begin{equation}
\left(
    \frac{d{\bf a}_k}{d\tau}
    +ik {\bf a}_k\frac{d \omega}{d\tau}
  \right) e^{ik\omega}
-\sum_l J_{k-l}\frac{d{\bf a}_l}{d\tau}
=
{\bf b}_k-{\bf a}_k e^{ik\omega}
\,.
\label{eq:torfr}
\end{equation}
Finally, the unit stability eigenvalue along the loop tangent direction
 ${\bf v}(s,\tau)$ needs to be
eliminated by adding to
(\ref{eq:torfr})
either the constant shift $\omega$ constraint
(\ref{eq:fixshift}),
or the {phase condition} 
(\ref{eq:nonsp1}) . in the Fourier representation the phase condition is given by
 $\sum_k k {\bf a}_k^* \cdot \partial {\bf a}_k/\partial \tau =0$ .

The monotone decrease with $\tau$ of the functional ${\mathcal F}^2$, given
by (\ref{eq:fvnal}), guarantees that the solution of (\ref{eq:torfr})
approaches a fixed point which, provided that  ${\mathcal F}^2\to 0$,
is the Fourier representation of the target
invariant torus.

In our numerical calculations, we represent the loop by a
discrete set of points
$\{{\bf x}(s_1), \cdots, {\bf x}(s_{2N})\}$.
The search is initialized by a $2N$-point guess loop.
The Fourier transforms of ${\bf x}$, ${\bf v}$ and $J$ are computed
numerically, yielding $M$ complex Fourier coefficients ${\bf a}_k$, ${\bf
b}_k$, and $J_k$, respectively. To maintain numerical accuracy, we choose
$M \leq N$
and
set ${\bf a}_k=0$,
${\bf b}_k=0$, and $J_k=0$ for $|k| \geq M$.
We terminate the numerical integration of the fictitious time
dynamics (\ref{eq:expfrev}) when the
distance (\ref{distMap}) falls bellow a specified cutoff. In
the Fourier representation, we stop when distance
reaches the {\em termination value} $\Delta$ defined as
\begin{equation}
\max_k \Vert {\bf F}_k\Vert
=
\max_{k,j}   |b_{k,j}-a_{k,j} e^{i k\omega} |< \Delta
\,,
\label{eq:terminate}
\end{equation}
where $a_{k,j}$ and $b_{k,j}$ denote the $j$th component of 
${\bf a}_k$ and ${\bf b}_k$.

While the algorithm is more efficient the better the initial guess,  in 
practice it often works for rather inaccurate initial guesses.
If the initial guess is bad, or the target invariant torus does
not exist, the evolution diverges. Then
another search is initiated,  with a new guess. This guess torus can either 
be derived from the integrable limit, like the examples of
Secs.~\ref{sec:standm}, \ref{sec:forcP} and \ref{sec:2stMap}, or from numerical
exploration, like the example of Sec.~\ref{sec:KS}. If
the invariant torus is isolated or partially hyperbolic, it can be a 
challenging problem to initialize the search for an embedded invariant torus. However,
once provided with a reasonable guess, our method is able to reliably locate the 
torus with relatively high accuracy.

Another concern is related to the numerical efficiency. If we try to find a higher order
torus (large $m$) in a high dimensional phase space (large $d$) with high 
accuracy, we have to repeatedly invert a very large
$[((2M)^m d+m) \times ((2M)^m d+m)]$ matrix when carrying out the
integration of Eq.~(\ref{eq:torfr}). This may
constitute a major bottleneck in such calculations. 
In our numerical implementation, the matrix  invertsion by the LU 
decomposition~\cite{nr} consumes most of the computational time. 
We employ a
speed-up scheme, based on the continuity of the evolution 
of Eq.~(\ref{eq:torfr}). Once we have the LU decomposition at one step, we use it
to approximately invert the matrix in the next step, with accurate inversion 
achieved by iterative approximate inversion~\cite{nr}. In practice, we find that
one LU decomposition can be used for many $\delta \tau$ evolution steps. The more
steps we go, the more iterations at each step are needed to get the accurate 
inversion. After the number of such iterations exceeds some fixed given maximum 
number, another LU decomposition is performed. The number of integration steps
following one decomposition is an indication of the smoothness of the evolution, 
and we further accelerate our program by adjusting accordingly the step size
$\delta \tau$: the greater the number, the bigger the step size. Near the final 
stage of convergence, the evolution becomes so smooth that the step size can be
brought all the way up to $\delta \tau=1$, recovering the full undamped 
Newton-Raphson step and acquiring the desired quadratic
convergence.

\section{Examples}
\label{sec:num}

We now test the Newton descent method for determining invariant tori
on a series of systems of increasing dimensionality:
a two-dimensional area-preserving standard map, a
Hamiltonian flow with one and half degrees of freedom (a forced pendulum), a
4-dimensional symplectic map (two coupled standard maps), and a dissipative
PDE (the Kuramoto-Sivashinsky system). In the following, the representative 
points are uniformly distributed on the initial guess torus.

\subsection{Critical tori of the standard map}
\label{sec:standm}

As our first example we search for invariant 1-tori of a two-dimensional area-preserving map,
the standard map
 \begin{eqnarray}
        q_{n+1}  &=& q_n+p_{n+1}\,\qquad \mbox{ mod } 2\pi
        \nonumber\\
        p_{n+1} &=& p_n+K\sin q_n
\,, \label{eq:stmp}
\end{eqnarray}
 where $K$ is the nonlinearity parameter. For $K=0$ the map is a constant rotation in $q$,
and for $K>0$ its phase space is a mixture of KAM tori and chaotic regions.
In the Fourier space the initial guess loop ${\bf x}=(q,p)$ and
its image
${\bf f}({\bf x})=(q+p+K\sin q, p+K\sin q)$ are expanded as
\begin{eqnarray}
 {\bf x}(s,\tau)
    &=&{\bf s}+\sum_k {\bf a}_k(\tau) e^{iks}
        \,,\qquad {\bf s}=(s,0)
        \nonumber\\
 {\bf f}({\bf x}(s,\tau))
    &=& {\bf s}+\sum_k {\bf b}_k(\tau) e^{iks}
\,.
 \nonumber
\end{eqnarray}
The linear term ${\bf s}$ in Eq.(\ref{eq:torfr})  is needed 
to compensate the modulus $2\pi$
operation on $q$ in Eq.(\ref{eq:stmp}).
Substitution
into (\ref{eq:vtorus})
yields
 \begin{eqnarray}
&& \left(
      \frac{d{\bf a}_k}{d\tau}
       +ik {\bf a}_k  \frac{d\omega}{d\tau}
       \right) e^{ik\omega}
+\delta_{0\,k} \frac{d\omega}{d\tau}{\bf e}_1
-\sum_l J_{k-l}\frac{d{\bf a}_l}{d\tau}
        \nonumber\\ &&\qquad\qquad
              \,=\,
{\bf b}_k-{\bf a}_k e^{ik\omega}-\delta_{0\,k} \omega{\bf e}_1
\, , \label{eq:expfrev}
\end{eqnarray}
 where
 ${\bf e}_1=(1,0)$. If we denote by ${\bf F}_k$ the
distance (\ref{distMap}) on the right hand side of
(\ref{eq:expfrev}), the invariant torus condition
for constant shift (\ref{eq:fixshift}) is
${\bf F}_k={\bf 0}$ for all $k$, {\em i.e}
${\bf b}_k={\bf a}_k e^{ik\omega}$ for
$k\not=0$ and ${\bf b}_0={\bf a}_0+\omega {\bf e}_1$.

\begin{figure}[h]
\centering
\includegraphics[width=5.5cm]{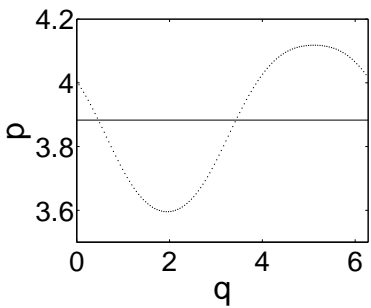}
\caption{
The
$\omega=\omega_g = 3.883\cdots$ golden mean invariant torus
of the standard map~(\ref{eq:stmp}) for
 $K=0.5$; the straight line represents the initial condition.
}
\label{f:sminv}
\end{figure}

 As the first test of our variational method, we apply
it to the determination of the golden-mean invariant torus,
with shift fixed to $\omega_g=2\pi(\sqrt{5}-1)/2$,
and the fixed shift constraint
(\ref{eq:fixshift}).
We use as the initial guess for the fictitious time dynamics
the invariant torus of the linear standard map
with $K=0$ and the golden-mean shift
$
{\bf x}(s,0) = (s,\omega_g)
$
,  represented by the straight line in Fig.\ref{f:sminv}.
In order to test that the method works for a smooth invariant torus we set
$K=0.5$ and integrate the fictitious time dynamics (\ref{eq:expfrev})
with
$2N=256$ point discretization of the torus,
$M=64$ complex Fourier mode truncation,
and $\Delta=2\times 10^{-6}$
termination value (\ref{eq:terminate}).
The resulting
invariant torus is shown by the dotted line in Fig.~\ref{f:sminv}.

Next, we apply the method to a
sequence of golden-mean invariant tori
with increasing $K$.
Numerics indicates that there exists a
critical value $\tilde{K}_c$ such that when $K<\tilde{K}_c$,
the fictitious time dynamics converges exponentially, as
in (\ref{eq:fvdc}),
but for $K>\tilde{K}_c$, it diverges. The critical value
$\tilde{K}_c$ depends sensitively on the torus discretization $2N$
and the termination value $\Delta$. $\tilde{K}_c(N)$ computed for
$\Delta=2\times 10^{-6}$ and several values of $N$ is
\begin{center}
\begin{tabular}{c|ccccc}
~$2N$ &               ~64&      ~128&    ~256 &     ~512 &     ~1024  \\ \hline
$\tilde{K}_c(N)$ &~~0.34 &~~0.80 &~~0.93 &~~0.9656 &~~0.9762   \\
\end{tabular}
\end{center}

\begin{figure}[h]
\centering
(a)~\includegraphics[width=5.5cm]{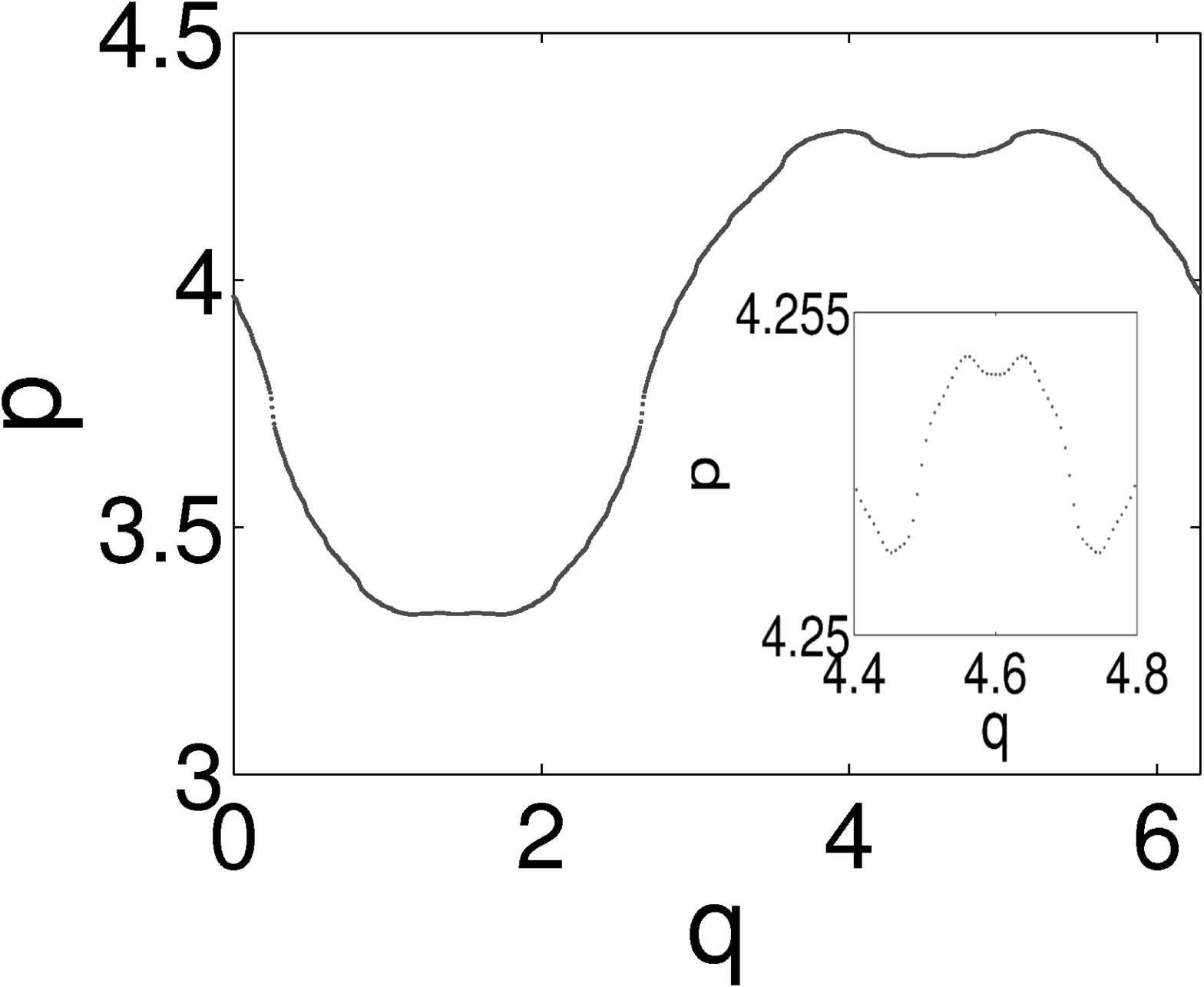}
(b)~\includegraphics[width=5.5cm]{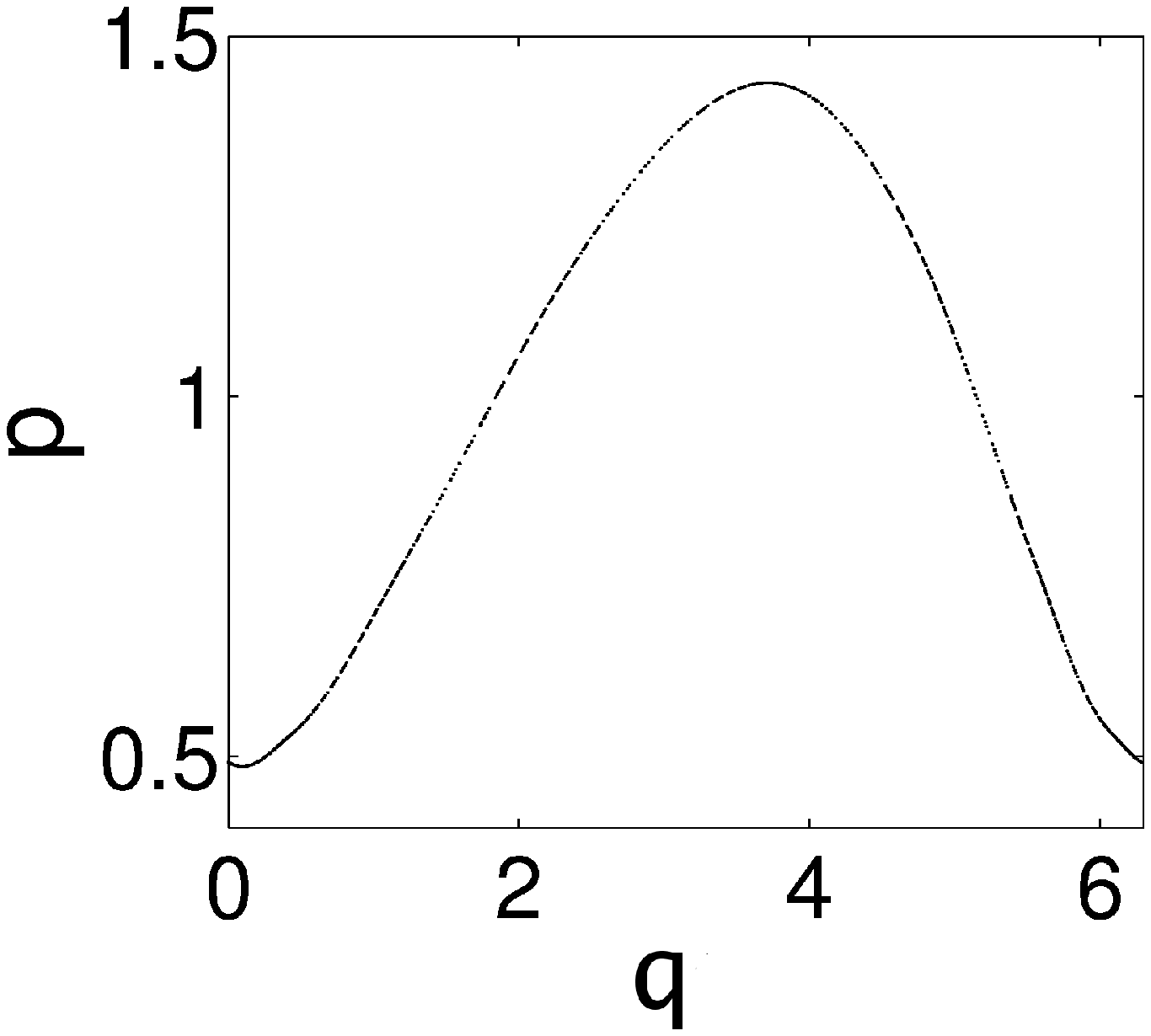}
\caption{
Invariant tori for the standard map~(\ref{eq:stmp}) for:
(a) $\omega=\omega_g$ at $K=\tilde{K}_c(512) = 0.9762$ close to the golden-mean torus
critical value $\tilde{K}_c$, termination value $\Delta=2\times 10^{-6}$.
The inset enlargement of the curve around $q=4.6$
illustrates the fine structure of the nearly critical torus.
(b) irrational shift $\omega=2\pi(\pi -3)$ at the estimated critical value $\tilde{K}_c(512)=0.4313$,
termination value $\Delta=4\times 10^{-6}$.
 $2N=1024$ torus points discretization.
    }
\label{f:torc}
\end{figure}

The golden-mean critical invariant torus is depicted in Fig.~\ref{f:torc}(a)
for $2N=1024$ points discretization of the torus.
Small oscillating structures in the critical torus whose resolution
would require higher frequency Fourier components are already
visible. The uneven distribution of representative points ($s$
parametrization's embedding into the $(q,p)$ plane) along the torus
indicates the drastically varying stretching rate on the invariant
torus close to the breakup~\cite{kada81,shen82}. Our variational
method of estimating the critical $K_c$ parameter is in agreement
with the Greene's estimate~\cite{gree79} that the golden-mean
invariant torus breaks up at the critical value $K_c\approx 0.9716 $.
 Moreover,
we find that for large values of $2N$ points discretization of the torus,
$\tilde{K}_c(N)$ approaches $K_c$ approximately as $N^{-1}$.

As Newton descent method does not depend on the specific
arithmetical properties  of the invariant torus shift, it should
work for arbitrary irrational shifts. As an example, we study the
family of invariant tori with shift $\omega=2\pi(\pi-3)$. The
critical value of convergence of our algorithm is
$\tilde{K}_c\approx 0.4313$ for $2N=1024$ and $\Delta=4 \times
10^{-6}$. The critical torus, depicted on Fig.~\ref{f:torc}(b)
exhibits non-uniform $s$-parametrization and oscillating structure,
though much less so than the golden-mean critical torus.

\begin{figure}[h]
\centering
(a)~\includegraphics[width=5.5cm]{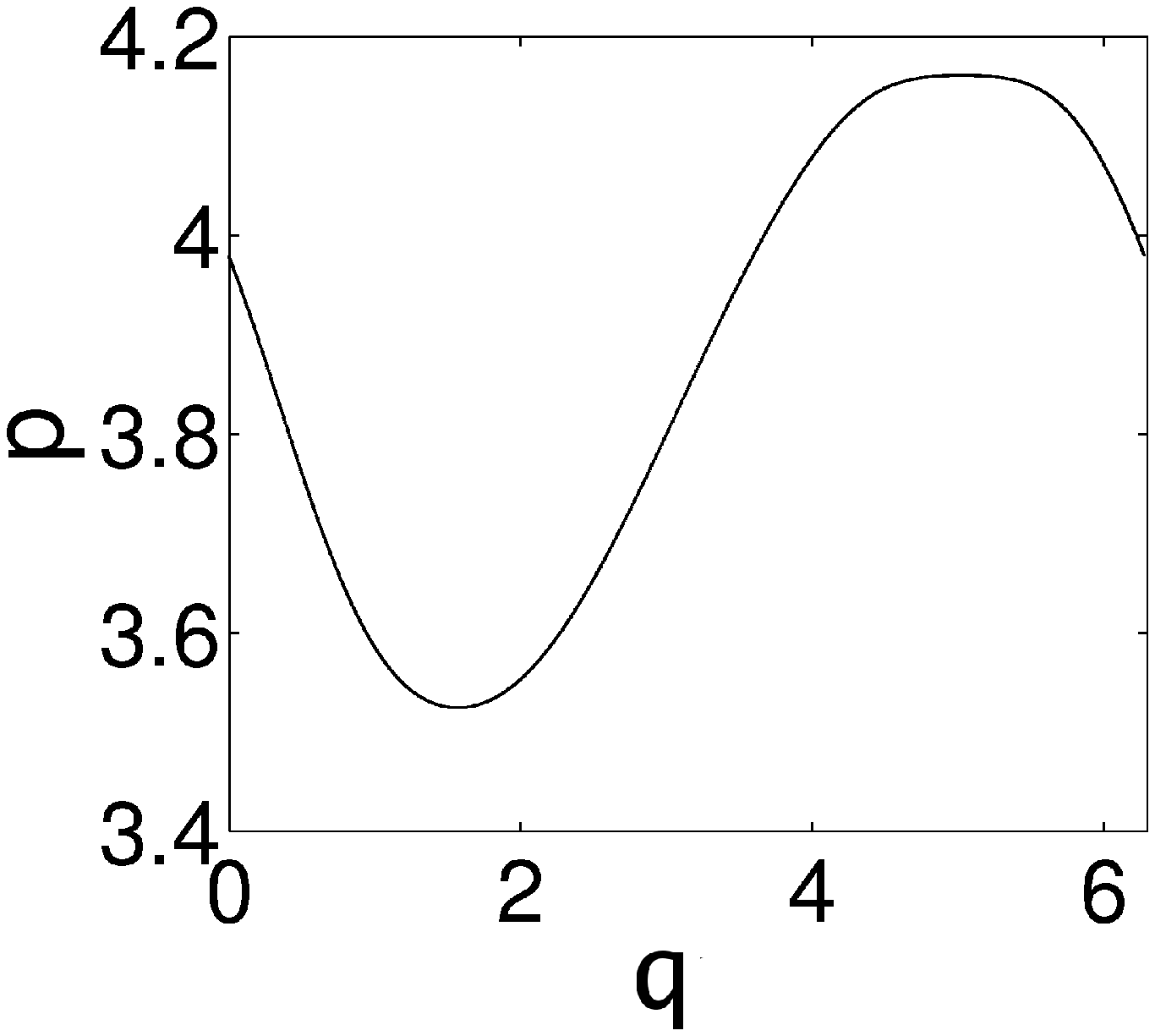}
(b)~\includegraphics[width=5.5cm]{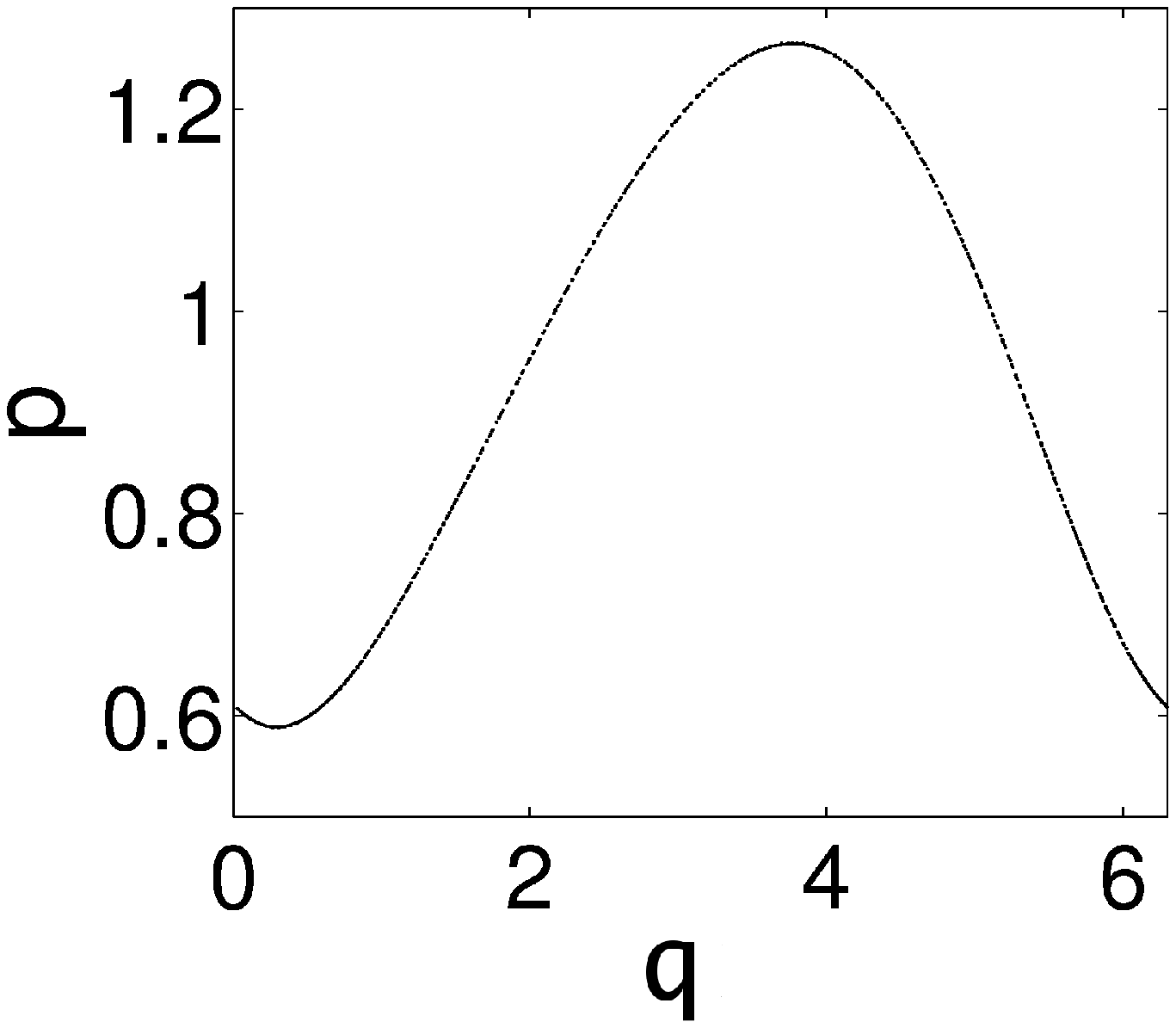}
 \caption{
 The invariant tori for the standard map
 (\ref{eq:stmp}) with smaller termination values $\Delta$ than in
Fig.~\ref{f:torc}, the same number of torus points $2N=1024$:
 (a)  $\omega=\omega_g$ with  $\tilde{K}_c =0.6188$ and $\Delta=10^{-6}$;
 and (b)  $\omega=2\pi(\pi -3)$ with $\tilde{K}_c=0.3004$
 and $\Delta=2\times 10^{-6}$.}
 \label{f:torcb}
\end{figure}

In order to assess the sensitivity of the method to the choice of
the termination value $\Delta$, we have studied its influence  on
the estimation of the critical $\tilde{K}_c$. For the golden-mean
example, a  decrease in the termination value to $\Delta=10^{-6}$
for $\omega=\omega_g$ and $2N=1024$ points discretization of the
torus, yields $\tilde{K}_c=0.6188$ much smaller than the value of
$\tilde{K}_c=0.9762$ obtained for $\Delta=2\times 10^{-6}$. The
corresponding invariant torus for $\Delta=10^{-6}$ is depicted in
Fig.~\ref{f:torcb}(a). We notice that this torus looks much smoother
than the one obtained for $\Delta=2\times 10^{-6}$ (see
Fig.~\ref{f:torc}(a)). Similarly, for $\omega=2\pi(\pi-3)$ a
decrease of the termination value to $\Delta=2\times 10^{-6}$,
yields a much smaller critical value $\tilde{K}_c=0.3004$. The
corresponding invariant torus for $\Delta=2\times 10^{-6}$ is shown
in Fig.~\ref{f:torcb}(b). The points are distributed more evenly
than in Fig.~\ref{f:torc}(b), indicating that the invariant torus
obtained using this termination value is far from criticality.

In summary: For fixed $2N$ points discretization of the torus,
 if $\Delta$ is too small, then $\tilde{K}_c(N) < K_c$, while if
$\Delta$ is too large, then $\tilde{K}_c(N) > K_c$. At the threshold
of criticality the invariant torus is fractal and thus cannot be
resolved by a smooth finite Fourier truncation.
The discrepancy between the invariant
torus and its numerical discretization has a complicated influence
on the fictitious time dynamics, not elucidated in this
investigation. If $\Delta$ is too small, the discrepancy leads to an
estimate of $\tilde{K}_c$ lower than the true $K_c$, making the
torus appear smoother.
If $\Delta$ is too large, the discretization will average out the
small features, converging to a grid beyond the critical value.
With increasingly refined $2N$ point-discretization of the torus, the
value of $\Delta$ needs to be chosen carefully in order to improve
the $K_c$  estimate.

\begin{figure}[h]
\centering
\includegraphics[width=6.0cm]{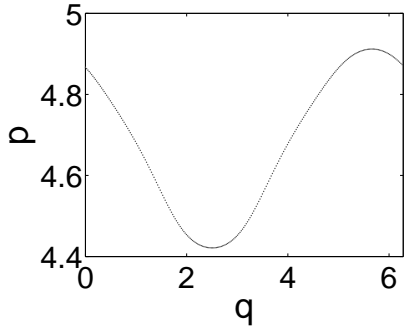}
\caption{
 An invariant torus of the standard map~(\ref{eq:stmp}) for $K=0.352$
 obtained by the fictitious
time dynamics with the phase condition~(\ref{eq:nonsp1}). The method
yields shift $\omega \approx 4.67857$.  $2N=256$ points
discretization of the torus, termination value $\Delta=2\times
10^{-6}$. }
 \label{f:smomg}
\end{figure}

So far we have determined invariant tori of the standard map
by imposing a constant shift condition (\ref{eq:fixshift}).
An alternative is the phase condition~(\ref{eq:nonsp1})
which requires that the motion of representative points along the
torus during the fictitious time dynamics averages to zero.
In this case the shift $\omega$
is not fixed, but is determined by the fictitious time dynamics.
We test this condition by starting with an initial torus ${\bf x}(s)=(s,9\omega_g/10)$ discretized
on $2N=256$ points, with termination value $\Delta=2 \times 10^{-6}$.
For $K=0.352$ the
Newton descent method yields the invariant torus of the standard map
shown in  Fig.~\ref{f:smomg}, with shift  $\omega\approx 4.67857$.

\subsection{A periodically forced Hamiltonian system}
\label{sec:forcP}

As our second test case, we
consider the forced pendulum
 \begin{equation}
H(p,x,t)={p^2}/{2}-\varepsilon (\cos x+\cos(x-t))
\,,
\label{eq:exph1}
\end{equation}
a time-dependent Hamiltonian flow with 1.5~degrees of freedom.
$H(p,x,t)$ is a periodic function of the angle
variable $x$ and the time variable $t$,
with
dynamics on $\mathbb{R}\times\mathbb{T}^2$.
The Poincar\'{e} return map for
the stroboscopic section $t=0\;\; \mathrm{mod~} 2\pi$
is a reversible area-preserving map.
The Jacobian $J$
required for the fictitious time dynamics (\ref{eq:vtorus})
is evaluated by integrating
\begin{equation}
\dot{J}=AJ\,, \;
A=\left(\begin{array}{lr}
0  &   1\\
-\epsilon (\cos x +\cos (x-t)) & 0
\end{array}\right)
\,, \;
 J(0)=1
\,. \label{eq:expha}
\end{equation}
 We apply the fixed shift condition
(\ref{eq:fixshift}) Newton descent to the determination
of the invariant torus with the golden-mean shift
$\omega=\bar{\omega}_g = (\sqrt{5}-1)/2$.
For the initial guess torus we take the golden-mean torus of Hamiltonian~(\ref{eq:exph1}) with $\varepsilon=0$, i.e.\ ${\bf x}(s)=(s,\bar{\omega}_g)$.
We define $\tilde{\varepsilon}_c(N)$ to be the minimum value of the parameter of the model at which
the algorithm defining the fictitious time dynamics with $2N$ sampling points fails to converge at fixed $\Delta$.
The critical values $\tilde{\varepsilon}_c(N)$ computed for different numbers of sampling points (termination
value $\Delta=2\times 10^{-6}$) are

\begin{figure}[h]
\centering
(a)~\includegraphics[width=5.5cm]{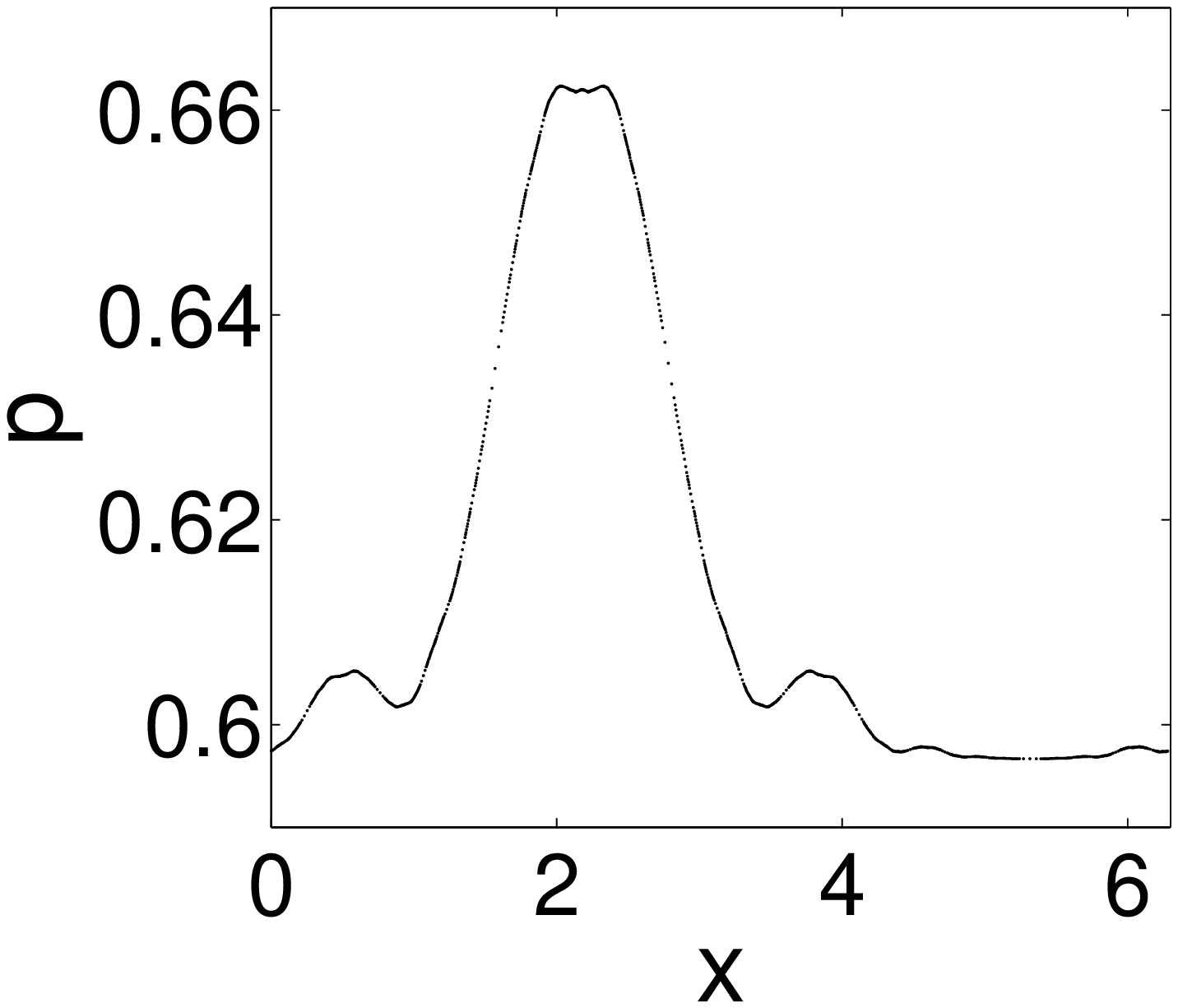}
(b)~\includegraphics[width=5.5cm]{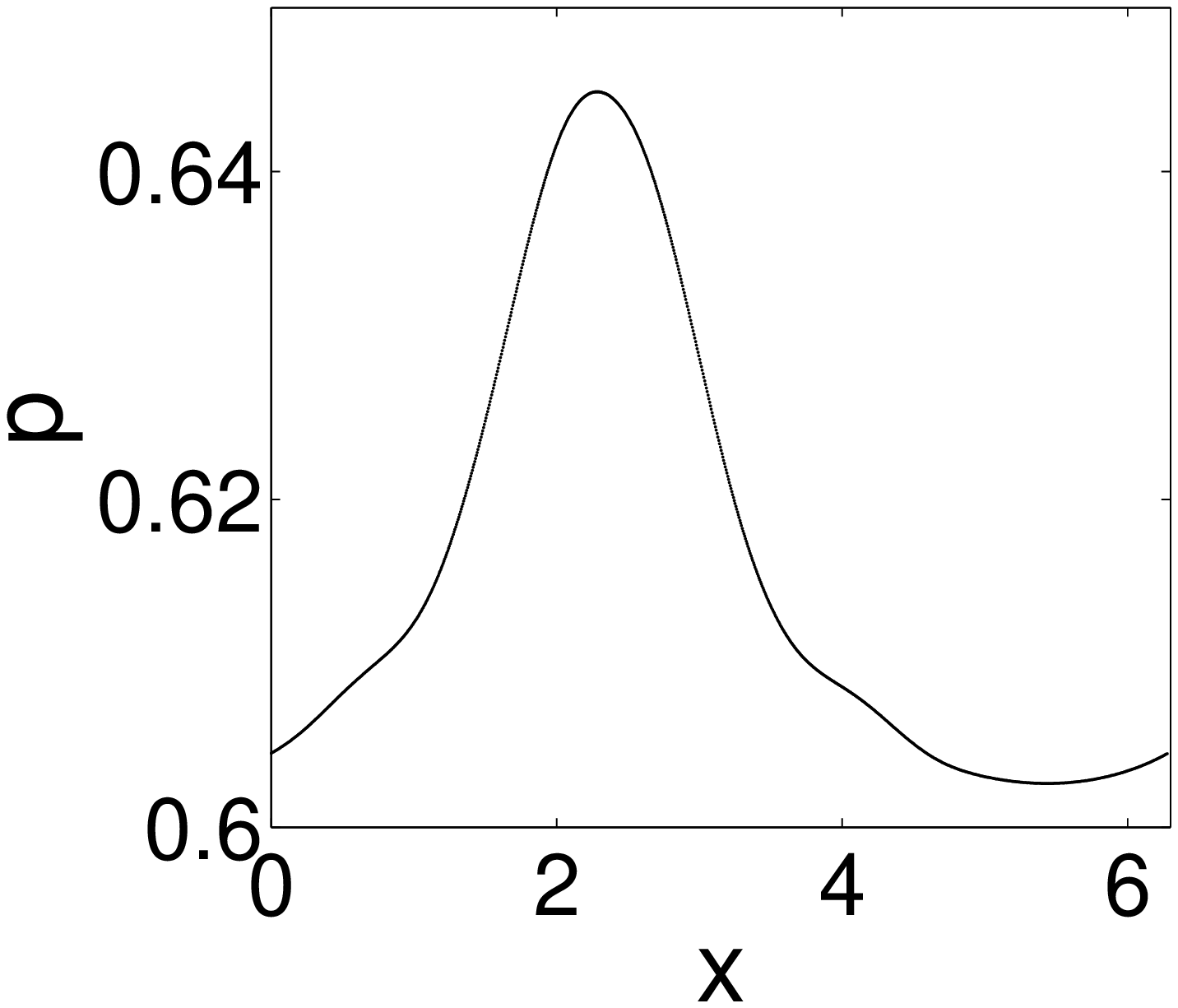}
\caption{Invariant tori of Hamiltonian~(\ref{eq:exph1})
with $\omega=\bar{\omega}_g$ obtained by the fictitious time dynamics
with $2N=1024$ and two different termination values:
(a) $\Delta=2 \times 10^{-6}$ yields a critical value
$\tilde{\varepsilon}_c=0.02781$, and
(b) $\Delta=10^{-6}$ yield to an underestimate
$\tilde{\varepsilon}_c=0.01844$.
 }
\label{f:khc}
\end{figure}

\begin{center}
\begin{tabular}{l|ccccc}
$2N$ &~64& ~128& ~256 & ~512 & ~1024 \\ \hline
$\tilde{\varepsilon}_c$ & ~~0.01688 & ~~0.02312 & ~~0.02594 & ~~0.02750 &  ~~0.02781 \\
\end{tabular}
\end{center}
 For $2N=512$ and $2N=1024$ the  $\tilde{\varepsilon}_c(N)$ values
 that we find are are close to the threshold
 $\varepsilon_c\approx 0.02759$ estimated in Ref.~\cite{PRchan}.
 The invariant torus with $\varepsilon=0.02781$,
$2N=1024$ and $\Delta=2 \times 10^{-6}$ shown in
 Fig.~\ref{f:khc}(a)
exhibits non-smoothness and an uneven distribution of
discretization points characteristic of criticality.
Setting $\Delta=10^{-6}$ leads to the
invariant torus with the critical value
estimate $\tilde{\varepsilon}_c=0.01844$, displayed in
Fig.~\ref{f:khc}(b).
It looks smooth, indicating that it is far from criticality
and thus that the termination value is too small.

\subsection{Two coupled standard maps}
\label{sec:2stMap}

In principle, the Newton descent method is applicable to
determination of invariant
tori of arbitrary dimension for flows or maps
of arbitrary dimension. In practice, one is severely limited
by computational constraints.

In order to test the feasibility of the method in higher dimensions,
here we consider two coupled standard maps~\cite{kan85arn},
 \begin{eqnarray}
I_{n+1}&=&I_n+\epsilon_1 \sin \theta_n +\epsilon_3 \sin (\theta_n+\psi_n)
    \nonumber   \\
\theta_{n+1}&=&\theta_n+I_{n+1}
    \label{eq:csm}  \\
J_{n+1}&=&J_n+\epsilon_2 \sin \psi_n +\epsilon_3 \sin (\theta_n+\psi_n)
    \nonumber   \\
\psi_{n+1}&=&\psi_n+J_{n+1}
\,, \nonumber
\end{eqnarray}
with 4-dimensional phase space,
and demonstrate that the method can determine
 1- and 2-dimensional invariant tori.
The fictitious time dynamics (\ref{eq:torfr})
acts on the ${\bf x}=(\theta_n,I_n,\Psi_n,J_n)$ phase space,
with dynamics ${\bf f}({\bf x})$ defined by (\ref{eq:csm}).

First, we apply the fixed shift (\ref{eq:expfrev})
fictitious time dynamics to determination of the
1-dimensional golden mean invariant torus with shift $\omega=\omega_g$.
For the initial guess torus we take the integrable case torus $\epsilon_1=\epsilon_2=\epsilon_3=0$~:
 \begin{equation}
 {\bf x}(s)=(s,\omega_g, s, \omega_g).
\label{eq:csminit1}
\end{equation}
In the numerical experiment we then
search for a typical 1-$d$ invariant torus, for
(arbitrarily chosen) small coupling values
$\epsilon_1=0.1$, $\epsilon_2=0.15$, $\epsilon_3=0.005$ .

\begin{figure}[h]
\centering
(a)~\includegraphics[width=5.5cm]{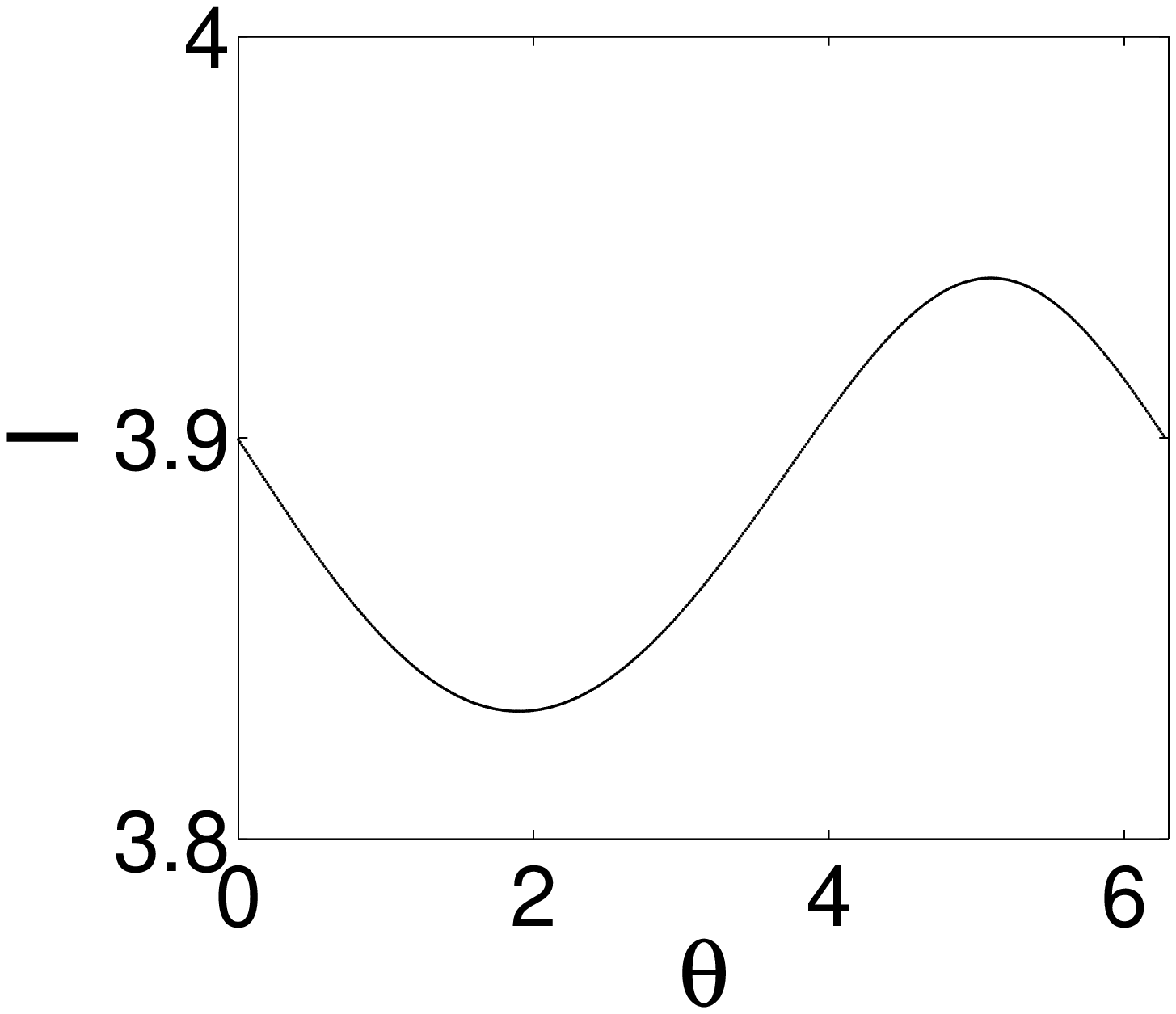}
(b)~\includegraphics[width=5.5cm]{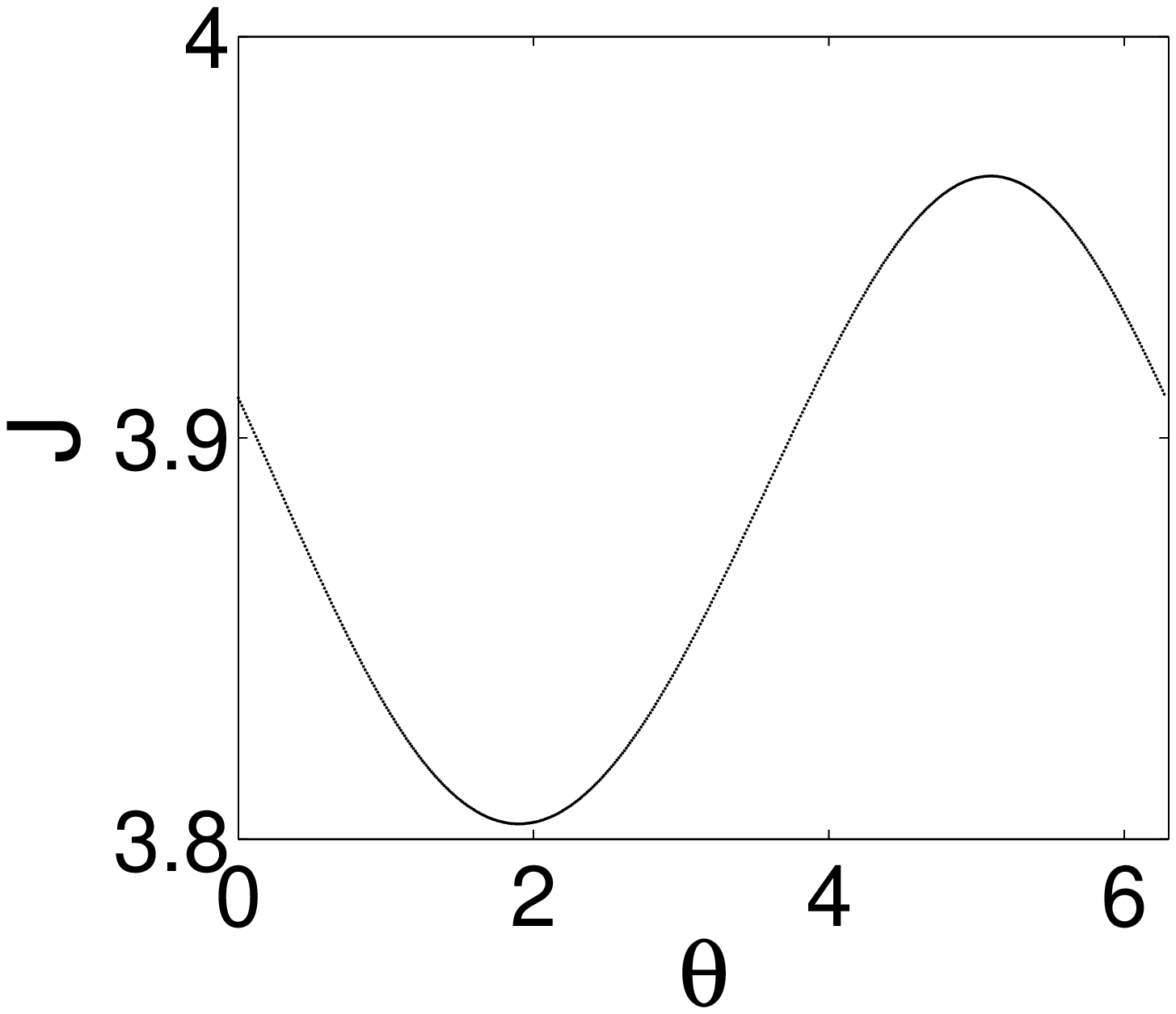}
\caption{ A 1-dimensional invariant torus with shift $\omega_g$ of (\ref{eq:csm})
with $\epsilon_1=0.1$, $\epsilon_2=0.15$ and $\epsilon_3=0.005$~: (a) $I-\theta$
projection;
(b) $J-\theta$ projection.
$2N=512$ points discretization of the torus,
termination value $\Delta=10^{-6}$. }
\label{f:tor4dt1}
\end{figure}

The invariant torus obtained by the fictitious time dynamics
in this case
is shown in Fig.~\ref{f:tor4dt1}.
Numerically $\theta=\psi$,
indicating that for this 1-dimensional torus the two phases are
entrained.
The torus appears very smooth, indicating that for the
parameter values chosen it is far from the critical values.

Next, we apply the Newton descent to
the determination of a 2-dimensional torus with
non-resonant frequencies $\omega_1$ and $\omega_2$. In this case, we need two
cyclic parameters $(s_1,s_2)\in [0,2\pi]^2$ to locate a
point on the torus. In (\ref{eq:expfrev}) we take
 \[
{\bf k}=\left( \begin{array}{c}
k_1 \\
k_2 \end{array}\right),\;
{\bm \omega}=\left( \begin{array}{c}
\omega_1 \\
\omega_2 \end{array}\right),\;
{\bf s}=\left( \begin{array}{c}
s_1 \\ 0 \\
s_2 \\ 0 \end{array}\right),\;
{\bf e}_1=\left( \begin{array}{cc}
1 & 0\\ 0 & 0 \\
0 & 1 \\0 & 0 \end{array}\right)
.
 \]
The initial guess is chosen as in the integrable
$\epsilon_i=0$ case
 \begin{equation}
 {\bf x}(s_1,s_2)=(s_1,\omega_1,s_2,\omega_2).
\label{eq:csminit2}
\end{equation}
In the numerical experiment we then search for
(arbitrarily chosen)
$\epsilon_1=0.07$, $\epsilon_2=0.1$ and $\epsilon_3=0.004$
2-dimensional invariant torus with
(also arbitrarily chosen)
frequencies
$\omega_1=\omega_g$ and $\omega_2=\pi(\sqrt{3}-1)$.
In order to reduce the
computational time, we take a rather coarse $2N=32$ grid,
with $(2N)^2=1024$ points representing the torus.

\begin{figure}[h]
\centering
(a)~\includegraphics[width=5.5cm]{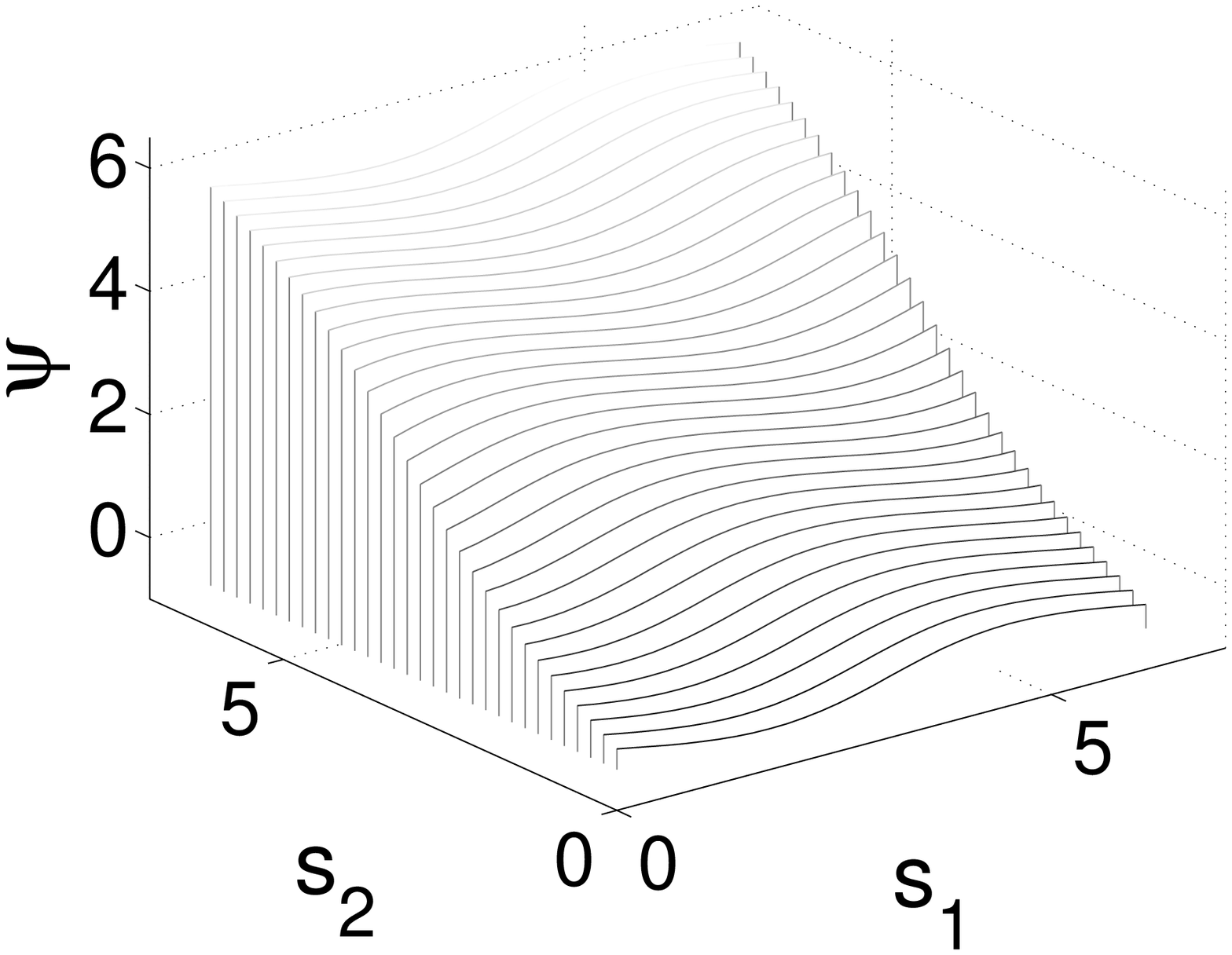}
(b)~\includegraphics[width=5.5cm]{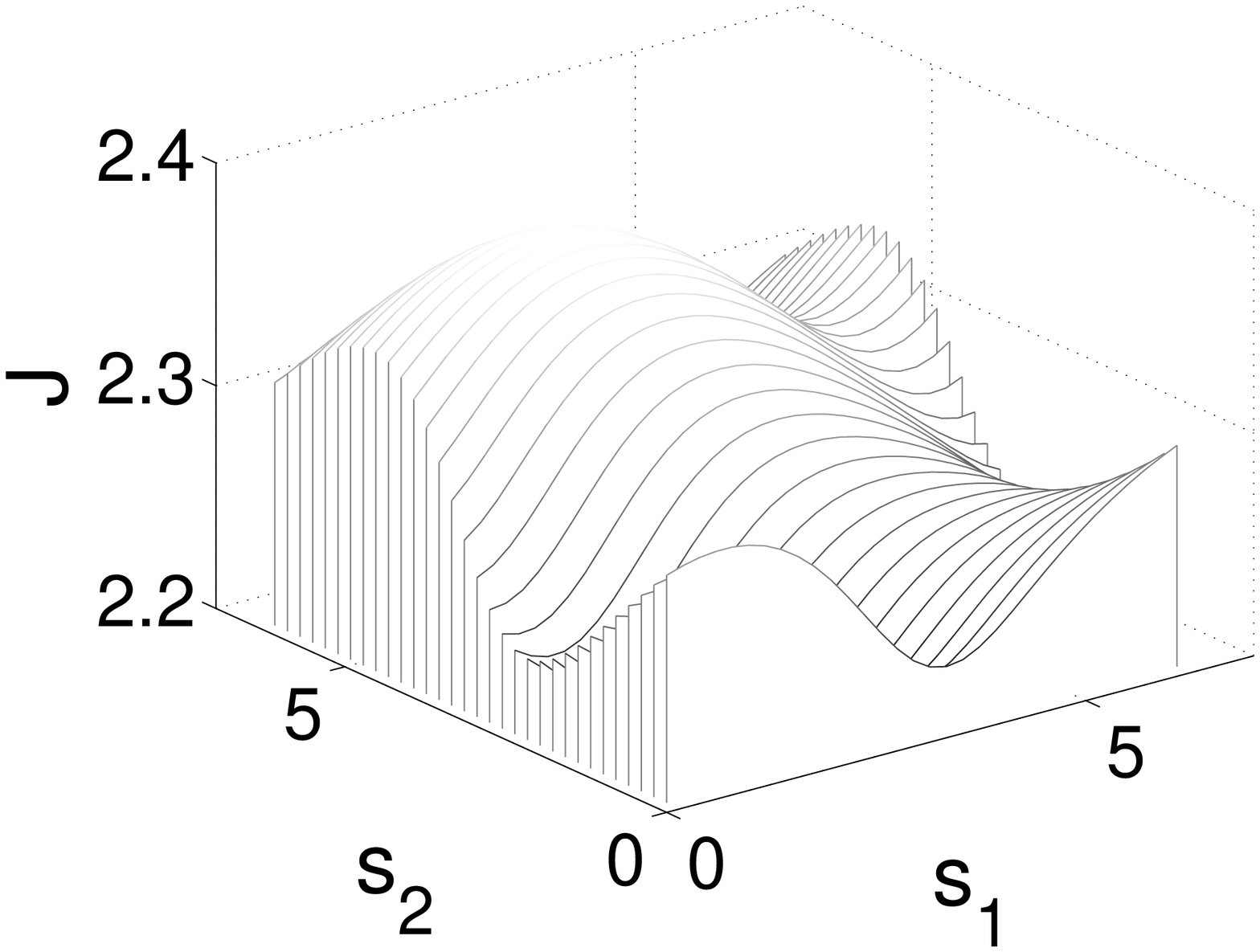}
\caption{
 The 2-dimensional invariant torus of the coupled standard
maps~(\ref{eq:csm}) with incommensurate
frequencies $\omega_1=\omega_g$ and $\omega_2=\pi(\sqrt{3}-1)$ for
$\epsilon_1=0.07$, $\epsilon_2=0.1$ and $\epsilon_3=0.004$.
$(2N)^2=1024$ points discretization of the torus, termination value
$\Delta=10^{-4}$.
    }
\label{f:tor4dt2}
\end{figure}

 Two projections of the resulting invariant torus
for $\Delta=10^{-4}$ termination value
are shown in Fig.~\ref{f:tor4dt2}.
While the $\psi(s_1,s_2)$ and $J(s_1,s_2)$
dependence on $s_1\,,s_2$
shown in Fig.~\ref{f:tor4dt2}
follows in shape the integrable case (\ref{eq:csminit2}) dependence,
the small coupling terms induce significant oscillations.
The smoothness of the invariant torus indicates
that the parameters are not close to the critical
values. For $(2N)^2=1024$ points discretization of the torus,
$\Delta$ can be as low as $5.1 \times 10^{-5}$, and
for $(2N)^2=4096$, as low as $1.6 \times 10^{-5}$.
However, the computation takes at least
$100$ times longer, and in this exploratory study the larger
$(2N)^2$ resolutions were out of reach.

\subsection{Kuramoto-Sivashinsky system}
\label{sec:KS}

In our last example, we apply the
Newton descent to determination of
an invariant 2-torus embedded in a high-dimensional strongly contracting
flow. 
Special tori that can be converted to periodic
orbits in a rotating or moving frame have previously been computed for the complex
Ginzburg-Landau equation~\cite{lop05rel}, and for the 2-d Poiseulle flow~\cite{cas00num}.
Here we shale determine a generic 2-torus of 
the Kuramoto-Sivashinsky equation~\cite{kuturb78,siv,HLBcoh98}
{parametrized by the system size $L$},
\begin{equation}
u_t=(u^2)_x-u_{xx}- u_{xxxx}
    \,,\qquad       x \in [0,L]
\,.
\label{eq:kseq}
\end{equation}
 The Kuramoto-Sivashinsky equation describes the interfacial
instabilities in a variety of contexts, like the flame front propagation,
the two fluid model and the liquid film on an inclined plane.

In the study of flame fluttering on a gas ring
as the
{system size $L$ increases}, the ``flame front'' becomes
increasingly unstable and turbulent.
As shown in Refs.~\cite{ruell71,nhouse78}, in dissipative systems
$2$-dimensional tori often result from a Hopf bifurcation
while $3$- (or higher-) dimensional tori are a rare occurrence.
 In the following we
restrict our search to the antisymmetric solution space of
(\ref{eq:kseq}) with periodic boundary conditions, i.e.\ $u(-x,t)=-u(x,t)$
and $u(x+L,t)=u(x,t)$,
with $u(x,t)$ Fourier-expanded as
 \begin{equation}
u(x,t)=\sum_{k=-\infty}^{\infty} i a_k e^{ikqx}, \label{eq:uexpan}
\end{equation}
 where $q=2\pi/L$ is the basic wavenumber and
 $a_{-k}=-a_k \in \mathbb{R}$.  Accordingly, (\ref{eq:kseq}) becomes
a set of ordinary differential equations~:
 \begin{equation}
\dot{a}_k=((kq)^2-
    (kq)^4)a_k - kq \sum_{m=-\infty}^{\infty} a_m a_{k-m}
\,.
\label{eq:ksexpan}
\end{equation}
 In the asymptotic regime of (\ref{eq:ksexpan}) for $k$ large
$a_k$'s decay faster than exponentially,
so a finite number
of $a_k$'s yields an accurate representation of
the long-time dynamics.  In
our calculation, a
truncation at $d=16$ suffices for a
quantitatively accurate calculation.

In the current example, $2N=128$ points are used to represent the torus on
the Poincar\'{e} section $a_1=0.06$.
Numerical experimentation indicates that for
$L=40.95$ trajectories spend significant fraction of time
in a toroidal neighborhood, suggesting that
a (partially hyperbolic?) invariant 2-torus exists at this
system size:
Poincar\'{e} section returns of a typical orbit
fall close to a closed curve.
The initial guess for the Newton descent is constructed by
choosing 128 points to represent this curve and their
Fourier transform is used to
initialize the search with (\ref{eq:expfrev}). In this case the shift
$\omega$ is fixed by dynamics, and in order to compute it
we impose the {phase condition}~(\ref{eq:nonsp1}).

\begin{figure}[h]
\centering
(a)~\includegraphics[width=5.5cm]{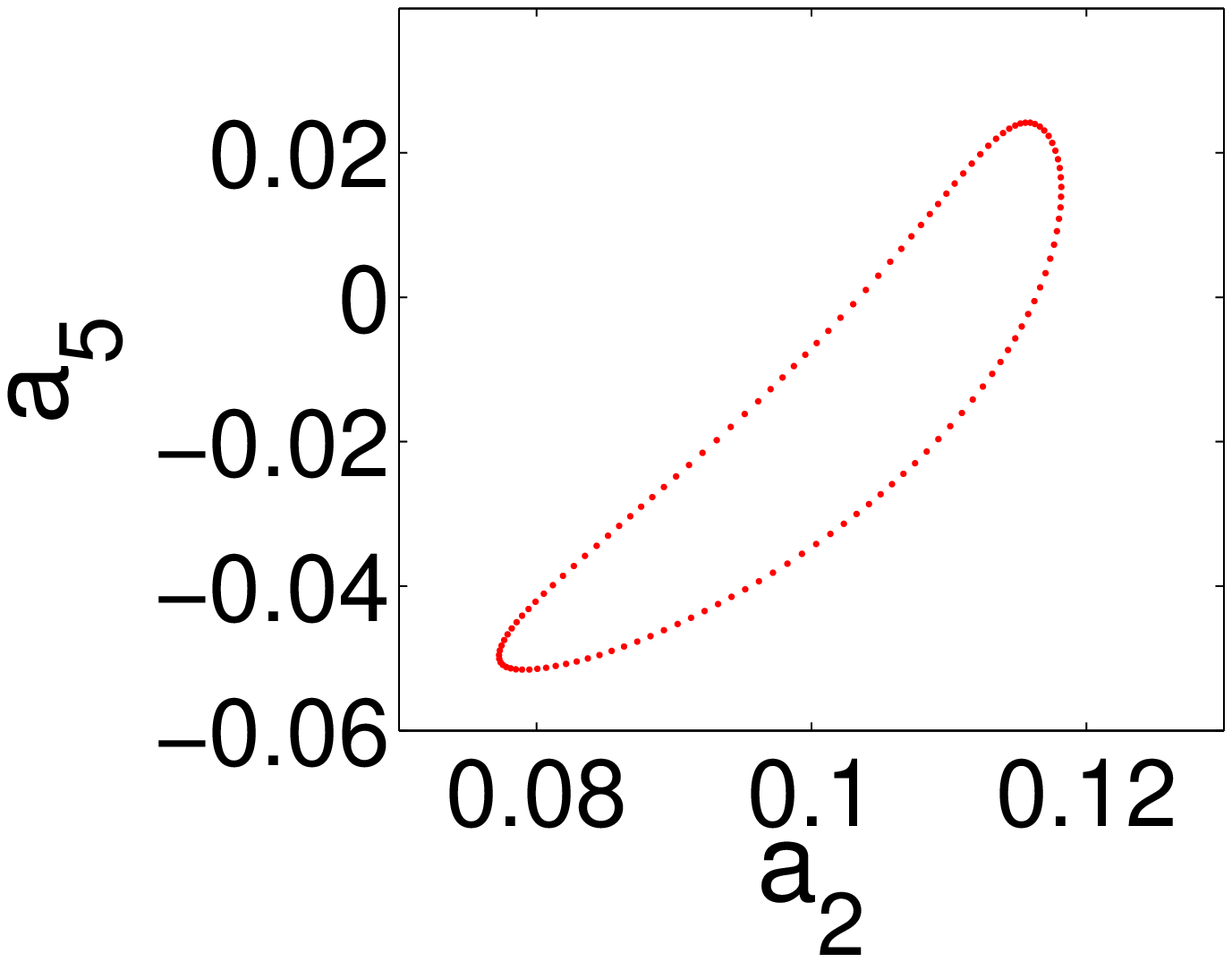}
(b)~\includegraphics[width=5.5cm]{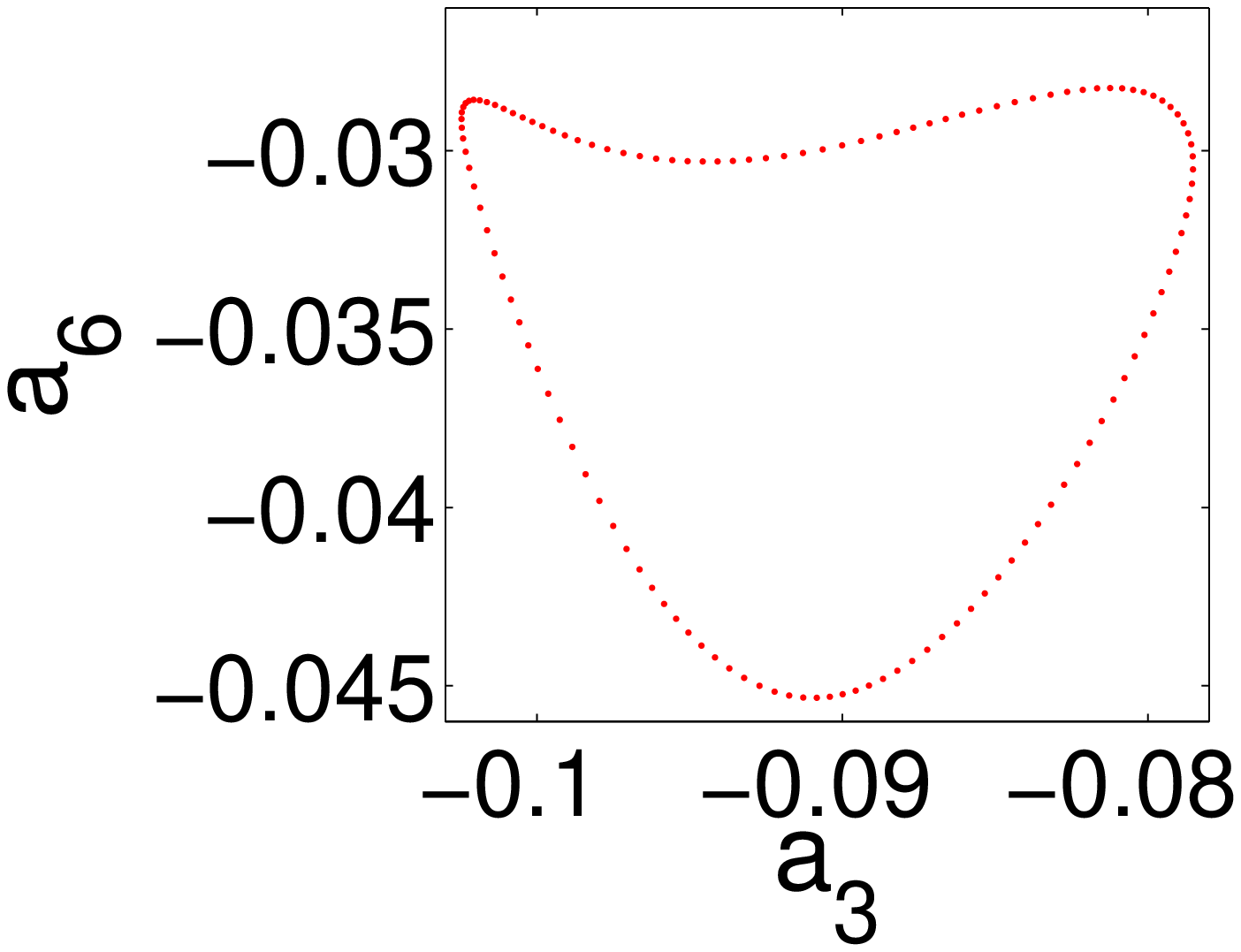}
\caption{
    The projections of the 2-dimensional invariant torus of
(\ref{eq:ksexpan}) on the Poincar\'{e} section $a_1=0.06$ with
shift $\omega=0.5968$ for
$L=40.95$~: Projection on (a)
$(a_2,a_5)$ and (b) $(a_3,a_6)$.
The Poincar\'e section return times are in the range
 $T=24.18 \pm 0.3$.
$2N=128$ torus points parametrization, $\Delta=10^{-4}$
termination value.
    }
\label{f:ks412a}
\end{figure}

Fig.~\ref{f:ks412a} shows two
Poincar\'e section projections, in the Fourier space,
 of the invariant
2-torus of the Kuramoto-Sivashinsky flow
determined by the Newton descent method. The method yields the
shift $\omega=0.5968$.
 Even though the invariant torus is very smooth and discretization
 points are evenly distributed,
 surprisingly many points are required to resolve the torus.
For attempts with fewer discretization points, for
example $2N=64$, the search did not converge even with
 $\Delta=10^{-2}$.

\section{Summary}

We have generalized the ``Newton descent'' variational method
to determination of invariant $m$-tori in
general $d$-dimensional dynamical systems,
and provided numerical evidence that the method
converges in a large domain of existence of invariant tori,
up to their breakups. In case of maps and flows with invariant tori
such as standard maps, the approach offers an alternative method
for determining critical thresholds.
While in principle the method is
applicable to flows or maps in arbitrary dimension, computation
can be expensive for invariant objects larger than 1- and 2-tori.
We have utilized the smoothness of the fictitious time 
evolution to introduce acceleration schemes which  improve the efficiency
of the method.

In our numerical work, we have implemented the method
in the constant shift (\ref{shiftMap}) parametrization,
Fourier representation of an $m$-torus.
Other discretizations  could be better suited to specific
applications. For
instance, if an invariant torus is close to its critical threshold,
representation of small fractal
structures requires inclusion of
slowly decaying high wavenumber Fourier modes,
and a large number of Fourier modes is needed to obtain an
accurate representation.
Furthermore, the discretization points
distribute very non-uniformly when close to criticality.
In this limit, other non-constant shift
parametrizations of the torus dynamics might
be more appropriate.
For example,
our method is of modest accuracy compared to some of current studies
of critical tori, in particular
Haro and de la Llave~\cite{delaLlave04:348}
computation of critical tori to 100 digits precision.

In periodic orbit searches we have found the Newton descent approach
robust, and very
useful for finding periodic orbits in high-dimensional phase-spaces where
good guesses for multi-shooting Newton routines are hard to
find~\cite{CvitLanCrete02,lanVar1}.
Examples worked out here suggest that the method is also a robust
starting point for $m$-dimensional invariant tori searches.
Once an approximate  invariant torus is found by the Newton descent method, it
can be used a starting guess for a high precision method,
such as some of the currently used Newton's methods in Fourier space
representations of invariant tori.

\end{document}